\documentclass[prb,aps,twocolumn,superscriptaddress]{revtex4-1}
\usepackage{graphicx,color}
\usepackage{amsthm}
\usepackage{amsfonts}
\usepackage{algorithmic}
\usepackage{enumerate}
\usepackage{latexsym}
\usepackage{amsmath}
\usepackage{amssymb}
\usepackage{bm}
\usepackage[colorlinks=true,citecolor=blue,linkcolor=blue]{hyperref}

\emergencystretch=\maxdimen
\begin{document}

\title{Tunable Dirac points and zero-energy modes in periodic curved graphene superlattices }% Force line breaks with \\

\author{Jianli Luan}
\affiliation{Department of Physics, Beijing Normal University, Beijing
100875, China\\}
\author{Kaiyi Guo}
\email{J. Luan and K. Guo contributed equally to this work.}
\affiliation{Department of Physics, Beijing Normal University, Beijing
100875, China\\}
\author{Shangyang Li}
\affiliation{Department of Physics, Beijing Normal University, Beijing
100875, China\\}
\author{Tianxing Ma}
\email{txma@bnu.edu.cn}
\affiliation{Department of Physics, Beijing Normal University, Beijing
100875, China\\}
\affiliation{Beijing Computational Science Research Center, Beijing 100193, China}
\author{Li-Gang Wang}
\email{sxwlg@yahoo.com}
\affiliation{Department of Physics, Zhejiang University, Hangzhou
310027, China\\}
\author{Hai-Qing Lin}
\affiliation{Beijing Computational Science Research Center, Beijing 100193, China}
\date{\today}
\affiliation{Department of Physics, Beijing Normal University, Beijing
100875, China\\}

\begin{abstract}
We combined periodic ripples and electrostatic potentials to form curved graphene superlattices and studied the effects of space-dependent Fermi velocity induced from curvature on their electronic properties. With equal periods and symmetric potentials, the Dirac points do not move, but their locations shift under asymmetric potentials. This shift can be tuned by curvature and potentials. Tunable extra gaps in band structures can appear with unequal periods. The existence of new Dirac points is proposed, such that these new Dirac points can appear under smaller potentials with curvature, and their locations can be changed even under a fixed potential by adjusting the curvature. Our results suggest that curvature provides a new possible dimension to tune the electronic properties in graphene superlattices and a platform to more easier study physics near new Dirac points.
\end{abstract}

\pacs{Valid PACS appear here}% PACS, the Physics and Astronomy
\maketitle

\section{\label{sec:level1}Introduction}
The Dirac fermion in curved two-dimensional space is an intriguing field of research due to its considerable applications in condensed matter physics\cite{Geim2007,PhysRevLett.103.196804,PhysRevLett.114.146803,PhysRevLett.121.221601}, materials science\cite{doi:10.1021/nl102661q,Martins_2010}, quantum field theory\cite{Birrel1982,Parker2009,PhysRevD.90.025006} and
astrophysics\cite{Hawking1976}. This topic has attracted interest for decades\cite{Birrel1982,Parker2009} but has received a particular boost due to the recent development of curved configuration of graphene superlattices (GSLs)\cite{Castro2009,Vozmediano2010,Rusponi2010,Yankowitz2012,Dubey2013,Ponomarenko2013,Ni2015,AMORIM2016}, in which the valance and conductance band touch at Dirac points (DPs)\cite{Castro2009}. We can also confine massless Dirac particles and measure a quantum spin-Hall current by implementing a curved Dirac equation solver based on the quantum lattice Boltzmann method.\cite{flouris2018confining,flouris2019quantum} Near the DPs in such curved graphene superlattices (CGSLs), energy bands have linear dispersion and electrons can be described by massless Dirac equations in a curved background. In addition to the CGSLs, there are various methods to induce different GSLs. In these structures, it has been predicted that the chirality of the charge carriers prevents the opening of a band gap and that, instead, new DPs appear\cite{Park20081,Park2009,AMORIM2016,Brey2009,Barbier2010,Wang2010,Dubey2013,Ponomarenko2013},
leading to a rich variety of remarkable results including the optical conductivity,
anisotropic Fermi velocity \cite{Park20081,Park2009,AMORIM2016,Brey2009,Barbier2010,Wang2010,Dubey2013,Ponomarenko2013,Park20082,Barbier2010,Yankowitz2012},
unconventional superconductivity \cite{Cao20181,*Cao20182}, surface plasmons \cite{Sunku1153,Ni2018,*Ni2015}, etc. It has also been predicted that the energy gap may be modulated by scalar potentials.\cite{low2011gaps} Curvature generates electrochemical potential, which in turn stabilizes the curvature.\cite{kim2008graphene}
% Quality control editor: Please be consistent in whether or not a space is used before a citation.
These novel properties not only stimulate the development of interesting physics in various fields but are also important for designing future electronic and optical devices.

In GSLs formed by periodic potentials, DPs and the associated band gaps are robust \cite{Wang2010},
while the group velocity near the original DP only changes in one direction \cite{Brey2009,Barbier2010}; the existence of new DPs is also limited by strict conditions, and their locations cannot be tuned with a fixed potential \cite{Park2009,Brey2009,Barbier2010}.
%These properties serve as obstacles and do need further studies to overcome them timely.
As a two-dimensional material, graphene exhibits intrinsic ripples to maintain its stability \cite{Fasolino2007}.
Recently, fabricating controllable periodic ripples \cite{Tapaszt2012,Parga2008,Bao2009,Yan2013,Ni2012,Park2016} has provided new insight for realizing tunable Dirac points and zero-energy modes in CGSLs, as the curvature may play various roles in tuning electronic structures \cite{Castro2009,Vozmediano2010,AMORIM2016},
in which the space-dependent Fermi velocity \cite{Fernando2007,Fernando2012} and pseudomagnetic field \cite{Guinea2008,Sancho2016,Levy2010,Kun2019} are the most prominent ones. The electronic transport can been studied through the condensed matter approach with non-equilibrium Green's function (NEGF) method and general relativistic approach under a pseudo-magnetic field.\cite{stegmann2016current}
Some more optical properties can also be tuned by curvature, such as optical conductivity \cite{Chaves2014}, surface plasmons \cite{Smirnova2016} and the Wolf effect \cite{Xu2018}.
It is natural to introduce curvature to adjust GSLs' properties due to its controllability and wide effects, as well as its intrinsic characteristics in real materials observed in experiments.

Inspired by these studies, we construct periodic curved graphene superlattices by applying electrostatic potentials on curved graphene and study their electronic band structures. As the space-dependent fermi velocity plays an important role in graphene, which can produce bound states\cite{ghosh2017bound}, control resonant tunneling\cite{lima2016controlling}, break the symmetry between the electron and hole minibands\cite{lima2015electronic}, influence the total conductance, electronic structure, the Fano factor, guided modes, localized current density and the Goos-Hanchen shift\cite{lima2016controlling,lima2018tuning,lima2015electronic,wang2013guided,lima2017engineering}, etc. Here, we mainly concentrated on the effects of space-dependent Fermi velocity on its band structure of the curved periodic super-lattices. Due to the realization of controllable curved graphene \cite{Tapaszt2012,Parga2008,Bao2009,Yan2013,Ni2012,Park2016} and GSLs \cite{Dubey2013,Rusponi2010,Ponomarenko2013,Yankowitz2012,Ni2015}, our proposal may be realizable experimentally.

CGSLs have two periodic structures (i.e., curved surface and potential). We first study the case of equal periods and symmetric potentials. The locations of original DPs are robust against curvature, but the band structures' slope near the original DP (i.e., Fermi velocity) in directions both along and perpendicular to the potential wells decreases. That structure brings characteristics of new DPs \cite{Brey2009,Barbier2010} into the original DPs. Nevertheless, the DPs can shift in asymmetric potentials and their displacements can be changed by adjusting curvature and potentials. Since locations of Dirac points can be changed under fixed potentials by adjusting curvature in addition to adjusting potentials only \cite{Wang2010}, it may be more easily to study physics near new Dirac points. Tunable extra gaps can also exist with larger periods of curved surfaces. In addition, zero-energy modes \cite{Brey2009,Barbier2010} and the existence of new DPs are studied systematically. New DPs can appear with a smaller potential due to the space-dependent Fermi velocity. Moreover, their locations are also tunable. With their prominent roles in tuning electronic properties, CGSLs have wide potential in applications.% or some research on new DPs and transport.

\section{\label{sec:level2}Model and method}
The Hamiltonian of the low-energy electronic state in monolayer flat graphene reads \cite{Wallace1947}
\begin{equation}
H_m=
\hbar v_f(k_x\sigma_x+k_y\sigma_y),\label{eqflatH}
\end{equation}
where $v_f=10^6$m/s is Fermi velocity, $\sigma_x$ and $\sigma_y$ are Pauli matrices, and $\vec{k}=(-i\frac{\partial}{\partial x},-i\frac{\partial}{\partial y})$ is the wave vector from the $K$ point.

In the following, we turn to curved case\cite{Arias2015,Atanasov2015,Pavel2017,Fernando2013,Pavel2017,Pacheco2014,Yang2015,Flouris2018}. There are two widely used theoretical approaches to model the Hamiltonian of curved graphene. One approach is quantum field theory in curved spacetime \cite{Fernando2007,Fernando2012,Chaves2014,Arias2015,Atanasov2015,Pavel2017}. This method rests on the basis that the low-energy electron in graphene can be described by a massless Dirac equation, and graphene is regarded as a continuum object. One can obtain the Hamiltonian by combining the Dirac equation and the metric of the curved surface \cite{Vozmediano2010}. In this way, topological defects \cite{Fernando2007} and helicoidal graphene \cite{Atanasov2015} are investigated, and the space-dependent Fermi velocity is derived \cite{Fernando2007,Fernando2012}. The other way is the tight-binding model that takes into account the hopping's change caused by displacements of carbon atoms and strain \cite{Isacsson2008,Guinea2008,Sancho2016}. It successfully predicts the pseudomagnetic field \cite{Guinea2008,Sancho2016}. However, both approaches have a weakness in roundly describing the curvature's effects, so other theories were developed, such as considering strain in the metric \cite{Fernando2012,Fernando2013,Pavel2017}, using discrete differential geometry \cite{Pacheco2014}, employing the metric in reciprocal space \cite{Yang2015} and rewriting the Dirac equation to include strain \cite{Flouris2018}.

The space-dependent Fermi velocity and pseudomagnetic fields are two main effects of curvature. Here, we mainly focus on the former's effects, so we use the Dirac equation in curved spacetime to obtain the Hamiltonian. This method does not need to consider ripples extending in the armchair or zigzag direction, which is a convenient approach to apply potentials and is in the view of experiments \cite{Atanasov2015}.

We rewrite the Dirac equation into a covariant form to obtain it in curved spacetime and the Hamiltonian with the method proposed by previous researchers \cite{Vozmediano2010}. In this paper, we consider a one-dimensional periodic curved surface, which refers to ripples that are only dependent on one coordinate and are written as $z=h(x)$. The metric is
\begin{equation}
g_{\mu\nu}=\text{diag}(1, -(1+g^2(x)), -1). \label{eqmetric}
\end{equation}
where $g(x)=\frac{dz}{dx}=h^{\prime}(x)$. It is worth noting that the flat space time metric is adjusted for 1D ripples in the z dimension with the resulting "space" metric here. After calculations described in Appendix A, we obtain the Hamiltonian of graphene in this shape as
\begin{equation}
H_c=-i\hbar v_f(\frac{\sigma_x\partial_x}{\sqrt{1+g^2(x)}}+\sigma_y\partial_y).\label{eqcurH}
\end{equation}

Comparing Eq. (\ref{eqflatH}) with (\ref{eqcurH}), we find that graphene gets space-dependent decreased Fermi velocity in $x$-direction. The Hamiltonian above is non-Hermitian. However, it still has the real and positive eigenvalues while obeying the space-time reflection (PT) symmetry\cite{bender1998real} althought we cannot obtain analytic solution. One may modify Eq.(4) into the Hermite form $H'=\sqrt{v_F(x)}\bm{\sigma \cdot p} \sqrt{v_F(x)}$ \cite{peres2009scattering}. However, in order to solve the problem within the frame of the original two cpmponent pseudospin wavefunctions $\psi_{A,B}$, we will solve the differential equations derived from Eq.(4) .

%Then, we substitute $z=h(x)=a_0\cos(\alpha x)$ and obtain Eq. ({\color{blue}2}). %(\ref{eqsinH}).

The form of Hamiltonian above has been mentioned in the previous studies\cite{Fernando2007,Chaves2014}.Next, we will systematically study the influence of the periodic potential field and the modulation of the DPs. Since we do not take into account the discreteness of the lattice and the strain induced by ripples and that the spin connection \cite{Vozmediano2010} is zero in the one-dimensional curved surface, our Hamiltonian cannot include the pseudomagnetic field in curved graphene.
However, we are only concerned about the Fermi velocity and low-energy electronic states here, and we consider that GSLs are also continuum objects and that their lengths are larger than the deformation of lattices. This method is also appropriate for any ripple directions and more convenient in considering applied potentials. In addition, the tight-binding method cannot reveal the Fermi velocity's variation \cite{Guinea2008}.
Although other modified methods can account for other effects and derive other forms of the Fermi velocity \cite{Fernando2012,Flouris2018}, they are only different from ours in specific values. Thus, our concise model can still reveal the influences of the renormalized Fermi velocity well and fits this paper's concerns.

In previous experiments, controllable curved graphene has been realized \cite{Tapaszt2012,Parga2008,Bao2009,Yan2013,Ni2012,Park2016} and can be described by one-dimensional sinusoidal functions well \cite{Bao2009}.
Hence, we use the one-dimensional function $z=h(x)=a_0\cos(\alpha x)$ to model curved graphene with $a_0$ and $\Lambda=\frac{2\pi}{\alpha}$ representing its amplitude and period, respectively. The Hamiltonian with curvature can be written as
\begin{equation}
\begin{aligned}
H_c&=-i\hbar v_f(\frac{\sigma_x\partial_x}{\sqrt{1+a_0^2\alpha^2 \sin^2(\alpha x)}}+\sigma_y\partial_y) \label{eqsinH}
\end{aligned}
\end{equation}
Here, we set $f(x)=\sqrt{1+a_0^2\alpha^2 \sin^2(\alpha x)}$ to represent the effect of curvature. The period of $f(x)$ is $T_f=\frac{\lambda}{2}$. When $a_0=0$, then $f(x)=1$, and Eq. (\ref{eqsinH}) degrades into Eq. (\ref{eqflatH}). Comparing Eqs. (\ref{eqflatH}) and (\ref{eqsinH}), we find that graphene obtains a space-dependent decreased Fermi velocity in the $x$-direction. %Detailed calculations of deriving the Hamiltonian can be found in the supplemental material (SM) I.

\begin{figure}[tb]
\centerline{\includegraphics[width=8.5cm]{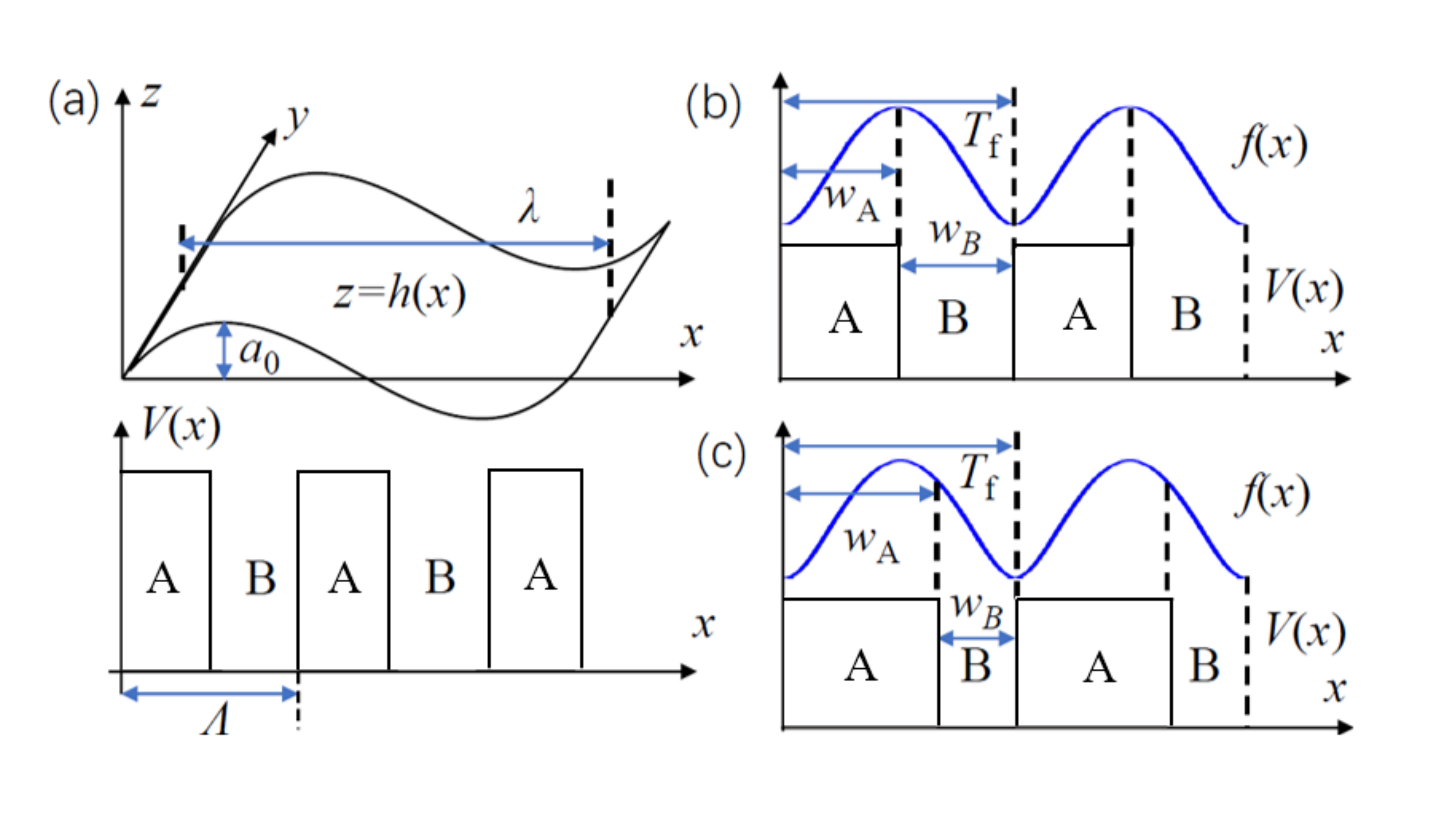}}
\caption{(Color online) (a) Schematic of one-dimensional sinusoidal curved graphene and periodic potentials with square barriers. (b) shows profiles of $f(x)$ and symmetric potentials. Here, $w_A=w_B$ and $T_f=\Lambda$. The $f(x)$ has the same values in the $A$ or $B$ regions. (c) shows the asymmetric case $w_A>w_B$, and $f(x)$ in $A$ has larger values than in the $B$ region. }
\label{model}
\end{figure}

Then, we apply one-dimensional periodic potentials $V(x)$ with square barriers on sinusoidal graphene and assume it is infinite in the $y$-direction to obtain a CGSL. The schematic of this CGSL is shown in Fig. \ref{model}. These potentials have the AB structure with a constant value inside each part, and widths are indicated by $w_A$ and $w_B$.
Thus, there are two periods in our model (i.e., $T_f$ and the potentials' period $\Lambda=w_A+w_B$); then, the lattice constant $T$ of a supercell in a CGSL is their lowest common multiple. Since scalar potentials in curved graphene can be added directly to the Hamiltonian \cite{Toyama1999}, we have the total Hamiltonian
\begin{equation}
H=H_c+V(x)I,\label{eqSL}
\end{equation}
with a 2$\times$2 unit matrix $I$. The Hamiltonian (\ref{eqSL})
acts on the two-component pseudospin wave function $\Psi=(\tilde{\psi}_A, \tilde{\psi}_B)^T$, and $\tilde{\psi}_{A,B}$ indicate smooth enveloping functions for the $A$ and $B$ sublattice in graphene \cite{Wang2010}. $\tilde{\psi}_{A,B}$ are written as $\psi_{A,B}e^{ik_yy}$ because of translation invariance.

To solve the eigenequation of $H$, we need to obtain the transfer matrix according to the construction by Ref.\cite{Wang2010}. The main problem of the model described by Eq. (\ref{eqsinH}) is that $f(x)$ is a continuous function, so it is not constant in the $m$th potential.
Therefore, we divided the $m$th potential into $n$ parts with an extremely small width.
Then, we can regard $f(x)$ as constant in each small part and use the value at the midpoint of the $j$th part $f_j$ to represent it. After this approximation, the transfer matrix that connects the wave function from $x$ to $x+\Delta x$ in the $j$th part reads
\begin{equation}
{M_j}(\Delta x,E,{k_y}) = \left( {\begin{array}{*{20}{c}}
{\frac{{\cos ({q_j}\Delta x - {\theta _j})}}{{\cos {\theta _j}}}}&{i\frac{{\sin ({q_j}\Delta x)}}{{\cos {\theta _j}}}}\\
{i\frac{{\sin ({q_j}\Delta x)}}{{\cos {\theta _j}}}}&{\frac{{\cos ({q_j}\Delta x + {\theta _j})}}{{\cos {\theta _j}}}}
\end{array}} \right). \label{TM}
\end{equation}
Although it has the same expression as flat graphene \cite{Wang2010}, the parameters in it are totally different and reflect the effects of potentials and curvature (see below). In Eq. (\ref{TM}), $\theta_j$ represents the incident angle of wave functions and $\sin\theta_j=\frac{k_y}{k_j}$, $\cos\theta_j=\frac{q_j}{f_jk_j}$ with the wave vector $k_j=\frac{E-V_j}{\hbar v_f}$ and satisfy the relation $(f_jk_j)^2=(q_j)^2+(f_jk_y)^2$.
%$q_j$ in Eq. (\ref{TM}) should be
 Hence, $q_j$ indicates $x$-component of wave vector and reads
\begin{equation}
q_j=
\begin{cases}
\text{sgn}(k_{j})\sqrt{(f_jk_j)^2-(f_jk_y)^2},\quad k_j^2>k_y^2\\
i\sqrt{(f_jk_y)^2-(f_jk_j)^2},\quad \text{otherwise}
\end{cases}.
\end{equation}
The $f_j$ reflects the influences of curvature, and Eq. (\ref{TM}) is also valid for the flat situation when $f_j=1$. For the case of $k_j=0$, Eq. (\ref{TM}) should be replaced by
\begin{equation}
M_j(\Delta x,E,k_y)=\text{diag}(e^{f_jk_y\Delta x}, e^{-f_jk_y\Delta x})
\end{equation}
The determinants of the above matrices both fit det$[M_j]=1$.
Detailed parameters and processes for deriving transfer matrix can also be found in Appendix B.
% According to Eq. (\ref{TM}), one can calculate electronic band structures of infinite systems.
Then, we can determine that the transfer matrix that connects the two terminals of the $m$th potential should be
\begin{equation}
M_{m}(w_{m},E,k_y)=\prod_{j=1}^{n}M_j(w_{j},E,k_y),
\end{equation}
where $n$ is the total number of divided small parts in the $m$th potential.

For an infinite CGSL system (AB)$^N$ with $N\rightarrow\infty$, the electronic dispersion at any incident angle can be calculated from Bloch's theorem
\begin{equation}
\cos(\beta_xT)=\frac{1}{2}\text{Tr}\prod_{i=1}^{T/\Lambda}(M_AM_B),\label{eqband}
\end{equation}
where $T$ is the lattice constant of CGSL, $\Lambda$ is the period of potentials and $\beta_x$ is the $x$-component of the Bloch wave vector of the whole system. This relation is influenced by the two periods $\Lambda$ and $T_f$. If there is a real solution of $\beta_x$, an electron or hole state will exist in the band structure; otherwise, the band structure will show an energy gap. According to this, we can obtain band structures and find the locations of DPs.

The transport properties for a finite superlattice $(AB)^N$ system can also be calculated by Eq. (\ref{TM}).
We obtain the electronic reflection and transmission amplitudes from the continuity of wave functions \cite{Wang2010} with the property of det[$M_j$]=1 as follows:
\begin{equation}
r(E,k_y)=\frac{x_{22}e^{i\theta_0}-x_{11}e^{i\theta_e}-x_{12}e^{i(\theta_0+\theta_e)}+x_{21}}
{x_{22}e^{-i\theta_0}+x_{11}e^{i\theta_e}-x_{12}e^{i(\theta_e-\theta_0)}-x_{21}},\label{eqr}
\end{equation}
\begin{equation}
t(E,k_y)=\frac{2\cos\theta_0}
{x_{22}e^{-i\theta_0}+x_{11}e^{i\theta_e}-x_{12}e^{i(\theta_e-\theta_0)}-x_{21}},\label{eqt}
\end{equation}
where $\theta_0$ and $\theta_e$ are the incident and exit angle through the superlattice, respectively. $x_{ij}$ are the elements of the entire transfer matrix $X=\prod_{m=1}^{N}M_m(w_m,E,k_y)=\prod_{m=1}^{N}\prod_{j=1}^{n}M_j(w_j,E,k_y)$. Then, the transmission probability reads $T=|t|^2$. These transport properties of finite systems are another reflection of band structures in infinite ones.

%\noindent
%\underline{\it Results and discussion}--

\section{\label{sec:level3}results and discussion}

In this section, we calculate the band structures of periodic CGSLs from the above model to discuss the effects of the space-dependent Fermi velocity induced by curvature. We focus on the location of DPs, effective Fermi velocity and appearance of new DPs. To make our results more realistic, we choose a realizable amplitude and period of curved graphene in the experiments. Moreover, there are two kinds of curved graphene. One of them has an amplitude and period of nanometer length \cite{Tapaszt2012}, and the other has amplitudes of 0.7 nm-30 nm and periods of 370 nm-5 $\mu$m \cite{Bao2009}. For convenience, we choose amplitude $a_0$ and period $\lambda$ that are in the same order of magnitude with the latter case (i.e., $a_0$ is of order 10 nm and $\lambda$ is of order 10$^2$ nm) in our calculation.

We first consider the case in which the potential and $f(x)$ have the same periods, which refers to $T_f=\Lambda$, and symmetric potentials with $w_A=w_B$. Here, we choose $V_A=50$ meV, $V_B=0$ and $w_A=w_B=30$ nm, so the periods of the curved surface and $f(x)$ are $\lambda=120$ nm and $T_f=\frac{\lambda}{2}=60$ nm, respectively.

\begin{figure}[tb]
  \centering
  \includegraphics[width=8cm]{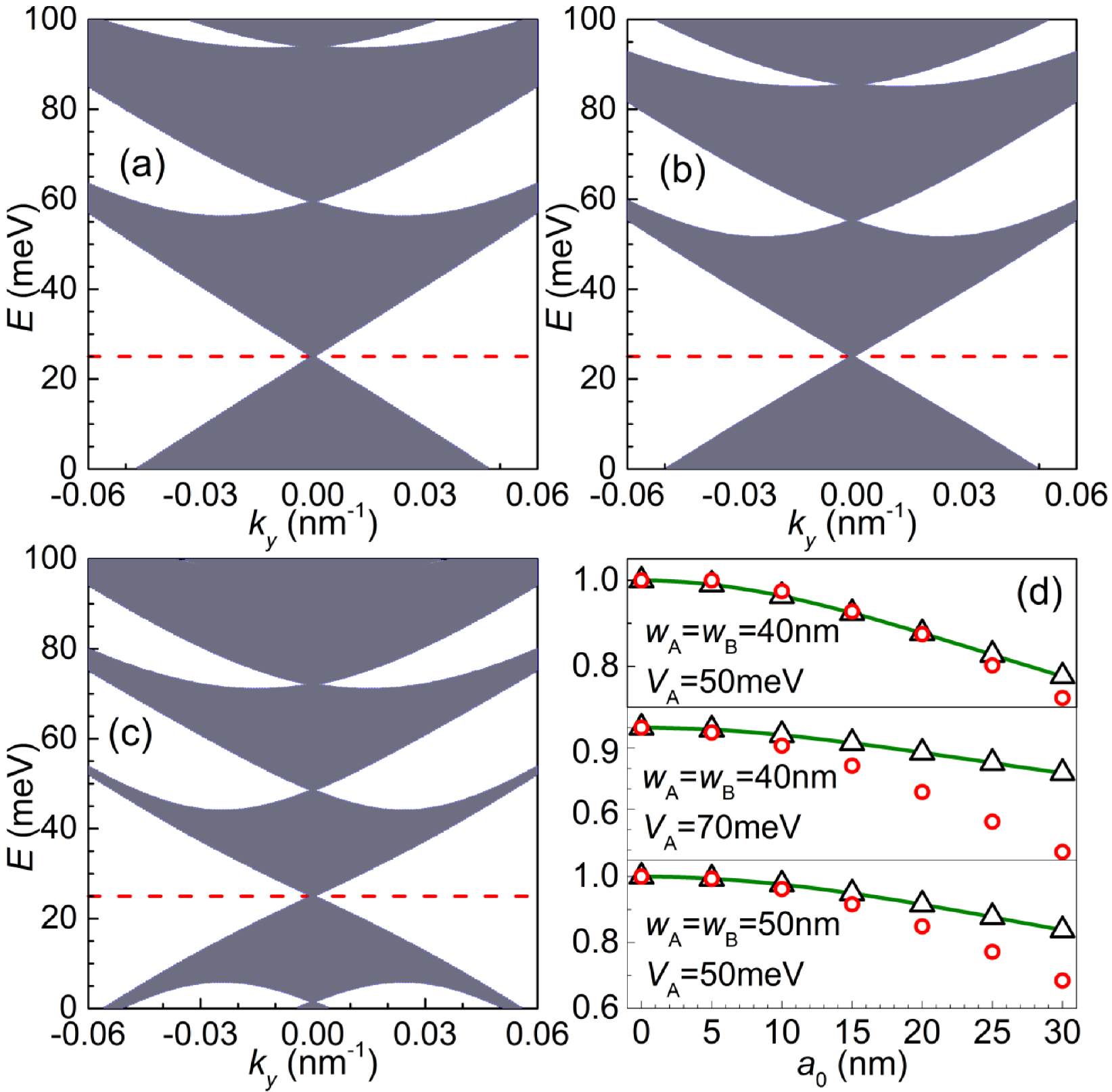}
  \caption{(Color online) Electronic band structures for $a_0=0$ (a), $a_0=15$ nm (b) and $a_0=30$ nm (c). Here, $V_A=50$ meV, $V_B=0$, $w_A=w_B=30$ nm and $T_f=\Lambda=w_A+w_B$. The red dashed lines denote the locations of DPs associated with the {\it zero-$\bar{k}$ gap}. The decreasing trend of the black triangle and red circle symbols in (d) denote the values of $\frac{v_x}{v_f}$ and $\frac{v_y}{v_{y0}}$, respectively, with different $w_{A,B}$ and $V_A$.
   Editor: Please ensure that the intended meaning has been maintained in the edits in the previous sentence.
   The green lines in (d) exhibit the value of $1/\overline{f(x)}$. Here, $V_B$=0 and $T_f=\Lambda$.}\label{figsymband}
\end{figure}

Figs. \ref{figsymband} (a),(b) and (c) show band structures of CGSLs with $a_0=0, 15, 30$ nm, respectively. From these bands, one can find that the locations of DPs associated with the {\it zero-$\bar{k}$} gap\cite{Wang2010} are 25 meV and robust with different $a_0$ values. This observation originates from the fact that potentials are symmetric and $\Lambda=T_f$, so $f(x)$ has the same values in the $A$ or $B$ potential, which is the case shown in Fig. \ref{model}(b) and is discussed in detail by a simplified model in SM III. However, with increasing $a_0$, the locations of other touching points of subbands with higher or lower energies are shifted and closer to the one at 25 meV. The widths of gaps associated with them and all subbands also decrease simultaneously.
This effect is due to the decreased slope of energy bands in the $k_y$-direction with increasing $a_0$. Meanwhile, the band structures' slopes in the $k_x$-direction also decrease obviously when $a_0\neq0$, despite being near the DP with the {\it zero-$\bar{k}$} gap. This property means that the effective Fermi velocity will decrease in both the $x$- and $y$-direction near original DPs in CGSLs and is completely different from that in flat GSLs, since near the original DPs in flat GSLs, the $x$-direction Fermi velocity is unchanged\cite{Barbier2010,Brey2009}.
%One can also find this property from the Hamiltonian Eq. (\ref{eqsinH}).
These results suggest that the space-dependent Fermi velocity induced by curvature works on band structures as a special potential and introduces the characters of new DPs\cite{Barbier2010,Brey2009} into CGSLs. Hence, CGSLs can be regarded as a platform to use new DPs' properties\cite{Luan2018}.

We also plot the ratio of the effective Fermi velocity between flat and curved GSLs in Fig. \ref{figsymband} (d). Here, we use $v_x$ and $v_y$ to indicate Fermi velocity in CGSLs. In flat GSLs, the $x$-direction Fermi velocity maintains $v_f$ and the $y$-direction  Fermi velocity decreases to $v_{y0}$. It is demonstrated that with increasing $a_0$ and fixed $T_f$, the ratios $\frac{v_x}{v_f}$ and $\frac{v_y}{v_{y0}}$ are both decreased, and $\frac{v_x}{v_f}$ is approximately equal to $\frac{1}{\overline{f(x)}}$. The $\frac{v_y}{v_{y0}}$ is nearly equal to it initially, but the distinctions between them increase dramatically with large $a_0$.
% Editor: Please ensure that the intended meaning has been maintained in the edits in the previous sentence.
These distinctions can originate from the variation of $f(x)$, which increases with $a_0$. The continuous variation of the curved surface can produce an effective potential, and it has been derived in helicoidal graphene \cite{Atanasov2015}.
Fig. \ref{figsymband} (d) suggests that the above relations are also robust with $V_{A,B}$ and $w_{A,B}$. Therefore, the influences of curvature can sometimes be represented partially by $\overline{f(x)}$.

\begin{figure}[tb]
  \centering
  \includegraphics[width=8.5cm]{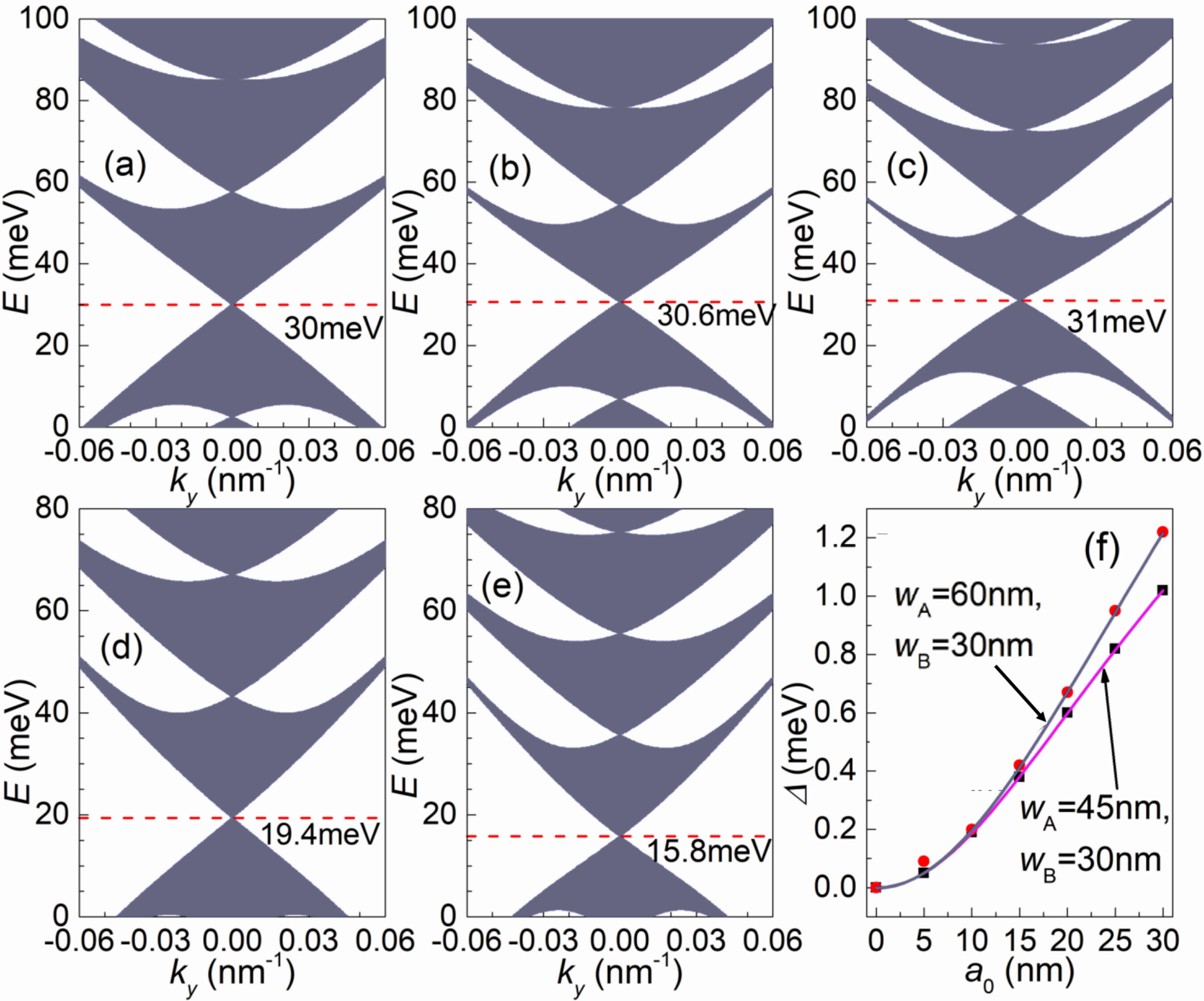}
  \caption{(Color online) (a)-(e) Band structures under asymmetric potentials. In (a)-(c), $w_A=45~nm$, $w_B=30~nm$, the amplitude of the curved surface $a_0=0$, 20 nm and 30 nm respectively. Here we set the value of $a_0$ to integers for convenience. In (d), $w_A=30~nm$, $w_B=45~nm$, $a_0=20$ nm and in (e), $w_A=30~nm$, $w_B=60~nm$, $a_0=24$ nm. Since the value of $\Lambda=w_A+w_B$ changed in (e), the value of $\alpha$, satisfied the formula $\Lambda=\frac{2\pi}{\alpha}$, also changed. We need to change the value of $a_0$ to $24~nm$ as well to fix the amplitude of $f(x)$, which is $a_0\alpha$. Other parameters are the same as in Fig. \ref{figsymband}(a). The red dashed lines denote the DPs' locations. The scattering diagrams in (f) are the shift of DPs with $w_A=45~nm$, $w_B=30~nm$ and $w_A=60~nm$, $w_B=30~nm$. The line shows the values calculated by Eq. (\ref{eqDPshift}) and the average of $f(x)$.}\label{figasymband}
\end{figure}

Then, we move to asymmetric potentials with $w_A\neq w_B$, and band structures are obtained in Fig. \ref{figasymband}. The most obvious feature in these figures is that DPs are shifted. From previous research, the DPs' locations in flat CGLs should be at 30 meV, 20 meV and 16.7 meV \cite{Wang2010} under the condition shown in Figs. \ref{figasymband} (a)-(c), (d) and (e), respectively. Thus, DPs move to higher energy when $w_A>w_B$ as in Figs. \ref{figasymband} (a)-(c). The opposite conclusion can be seen in Figs. \ref{figasymband} (d)(e) when $w_A<w_B$. This is because $f(x)$ has a larger value in the $A$ region than in the $B$ region with $w_A>w_B$ (shown in Fig. \ref{model}(c)) and is opposite that of $w_A<w_B$. Comparing Figs. \ref{figasymband} (b)(c) or (d)(e), one can also find that displacements of DPs increase with increasing $a_0$ and $w_{A(B)}$.

To understand the robustness and shift of DPs induced by the space-dependent Fermi velocity in different conditions, we propose a simplified theoretical model. We set $f(x)$ in each potential $A$ or $B$ as constant $f_{A(B)}$ and $T_f=\Lambda$. Since the average of $f(x)$ can reflect some effects of curvature, we set $f_{A(B)}$ as this average in the range of the potential. Then Eq. \ref{eqband} reduces to
%From Eq. (\ref{eqband}) and the assumptions proposed in the Results and discussions section, the band structures under this simplified model should be calculated by
\begin{equation}
\begin{aligned}
\cos(\beta_x\Lambda)=&\cos(q_Aw_A+q_Bw_B)\\
&+\frac{\cos(\theta_A-\theta_B)-1}{\cos\theta_A\cos\theta_B}\sin(q_Aw_A)\sin(q_Bw_B),\label{eqsimple}
\end{aligned}
\end{equation}
with $q_{A(B)}^2=(f_{A(B)}k_{A(B)})^2-(f_{A(B)}k_y)^2$, and the band structures under this simplified model should be calculated.
From the previous analysis, when $V_B<E<V_A$, the DPs in $k_y=0$ should exist and the locations are decided by $q_Aw_A=-q_Bw_B$ \cite{Wang2010}.
With $k_y=0$, it should be $f_Ak_Aw_A=-f_Bk_Bw_B$.
Substituting the expression of $k_{A(B)}$, we obtain the DP's location:
\begin{equation}
E=\frac{f_Aw_AV_A+f_Bw_BV_B}{f_Aw_A+f_Bw_B}.\label{eqDP}
\end{equation}
When $f_A=f_B=1$, Eq. (\ref{eqDP}) reduces to the flat situation. When $f_{A}\neq f_B$, the locations of DPs may shift,
and the displacement from flat situation $\Delta$ is shown by %Eq. (\ref{eqDPshift}). %({\color{blue}4}). %
%{\color{red} After the calculations shown in SM III} and comparing the case for $f_A=f_B=1$ with $f_{A}\neq f_B$, we can obtain the displacement of DP from the flat situation
\begin{equation}
\Delta=\frac{(f_A-f_B)(V_A-V_B)w_Aw_B}{(f_Aw_A+f_Bw_B)(w_A+w_B)}.\label{eqDPshift}
\end{equation}
In Fig. \ref{figsymband}, $f(x)$ has the same value in potential $A$ and $B$ since $w_A=w_B$, which corresponds to $f_A=f_B$ and $\Delta=0$.
In Figs. \ref{figasymband} (b) and (c), $f_A>f_B$ and $V_A>V_B$; then, $\Delta>0$ and DPs shift to higher energy. Figs. \ref{figasymband} (d) and (e) are the opposite. Meanwhile, Eq. (\ref{eqDPshift}) states that $\Delta$ is proportional to $w_{A,B}$, which is also consistent with the conclusion gained by comparing Figs. \ref{figasymband}(d) and (e).

% Editor: Please ensure that the intended meaning has been maintained in the edits in the previous sentence.

In Fig. \ref{figasymband}(f), the changes in the DPs' positions with different $a_0$ and $\Lambda$ values are demonstrated by the scattering diagrams. One can immediately find that with increasing $a_0$ and $\Lambda$, the DPs shift more, which agrees with Eq. (\ref{eqDPshift}) and suggests their tunability. We also computed the shift from the simplified model with $\overline{f(x)}$ and Eq. (\ref{eqDPshift}). These results are shown by the lines and fit the realistic results well. Therefore, constructing CGSLs can be a feasible way to tune the locations of DPs by adjusting both potentials and curved surfaces. Other ways that can change the Fermi velocity \cite{Pellegrino2012,Raoux2010,Krstaji2011} are also possible to tune DPs. Furthermore, one may need to concentrate on these effects in experiments since the intrinsic feature \cite{Fasolino2007} of ripples in graphene and potentials may not be strictly symmetric.

\begin{figure}[tb]
  \centering
 \includegraphics[width=8.5cm]{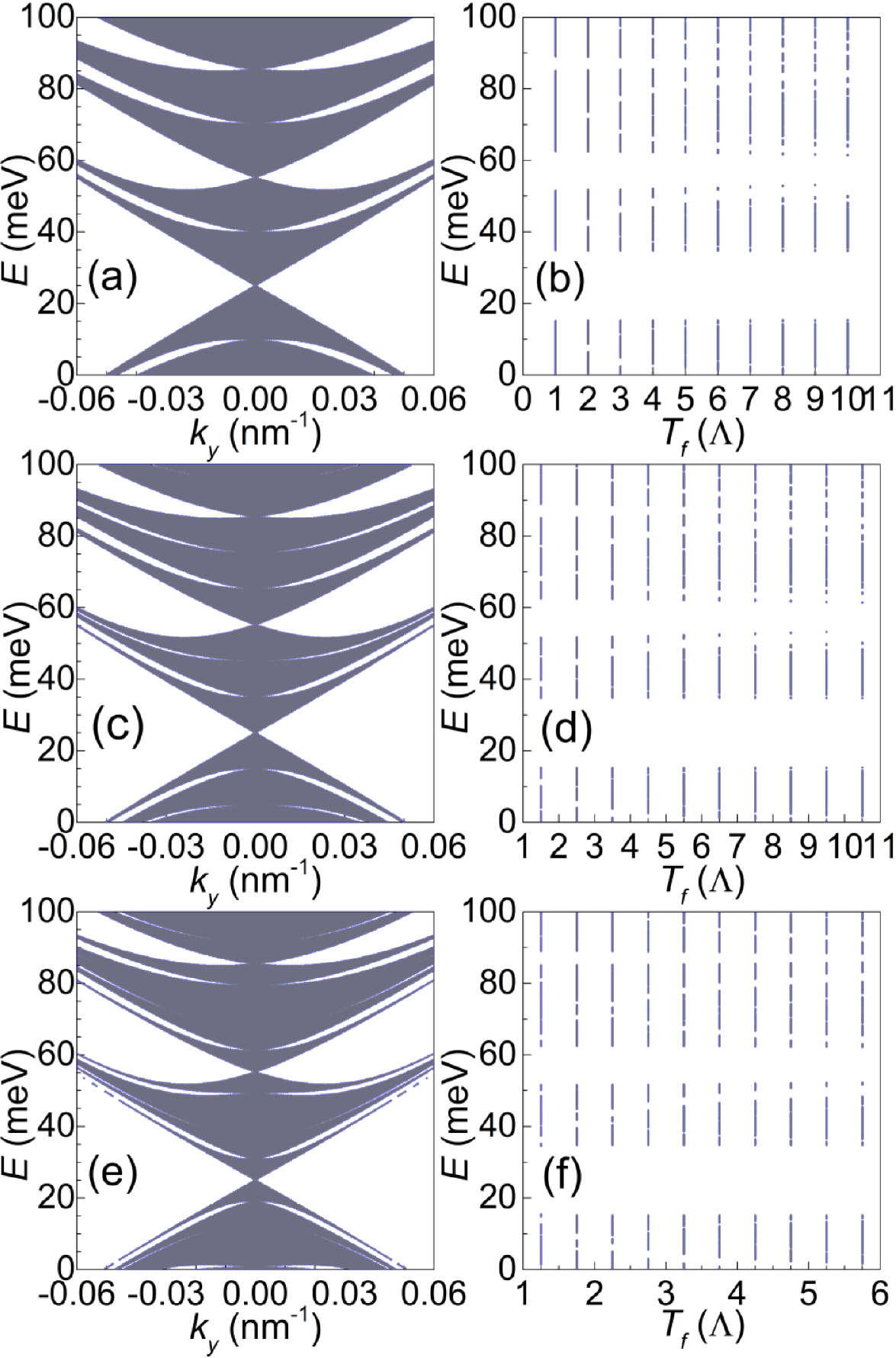}
  \caption{(Color online) Band structures when $T_f=2\Lambda$ (a), $1.5\Lambda$ (c) and $1.25\Lambda$ (e) with $a_0\alpha=\frac{\pi}{4}$. (b), (d) and (f) show the electronic states in different $T_f$ and $E$ with fixed $k_y=0.02$ nm$^{-1}$ and $a_0\alpha=\frac{\pi}{4}$. Other parameters are the same as in Fig. \ref{figsymband}(a). }\label{figgap}
\end{figure}

\begin{figure}[tb]
  \centering
  \includegraphics[width=8cm]{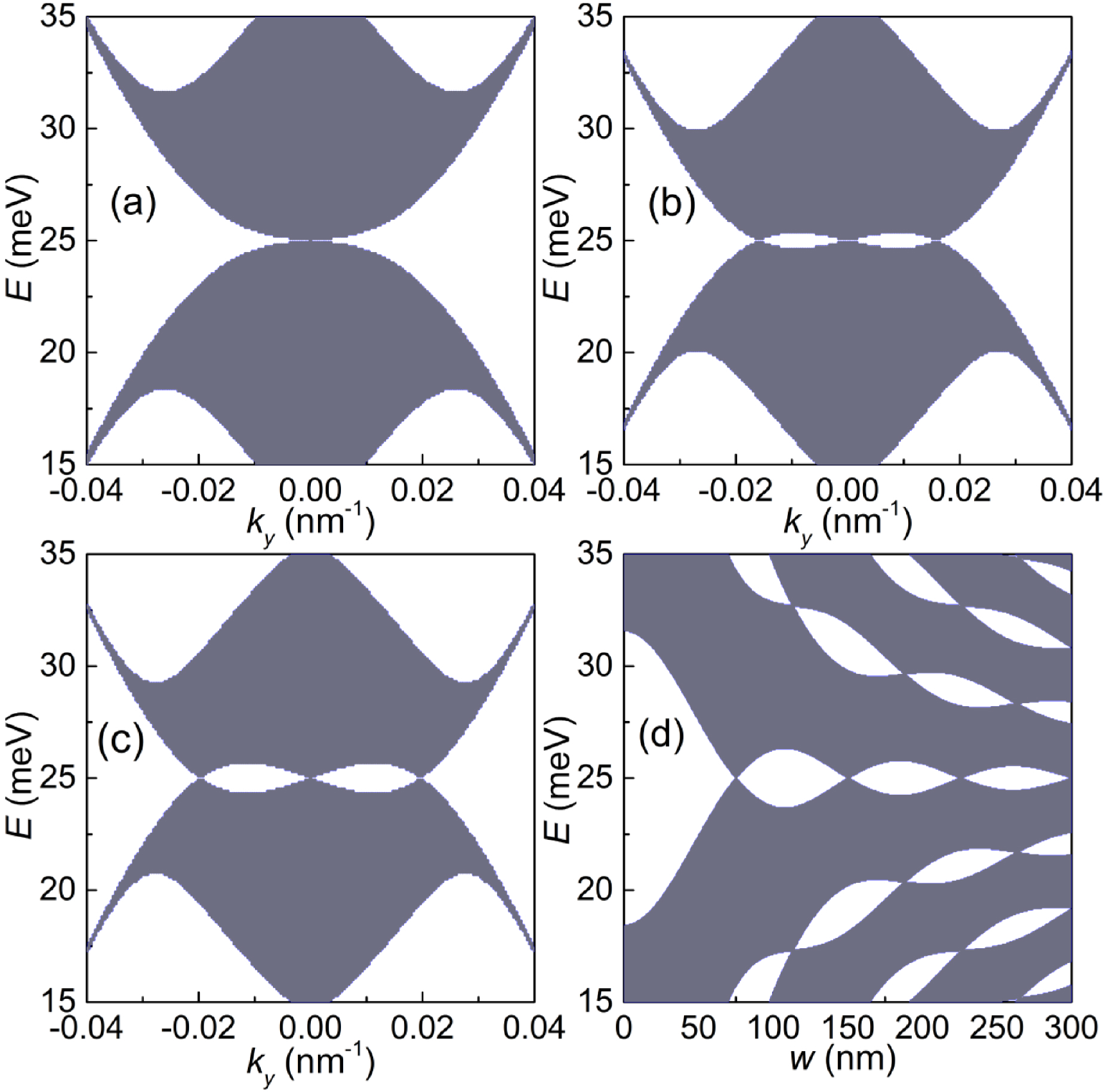}
  \caption{(Color online) Electronic band structures when $a_0=0$ (a), 40 nm (b) and 50 nm (c) with $w_A=w_B=80$ nm and $T_f=\Lambda$. (d) is the dependence of band structures on periods of potentials with fixed $k_y=0.01$ nm$^{-1}$ and $a_0\alpha=\frac{\pi}{4}$. Here, $w_A=w_B=w$ and $T_f=\Lambda=2w$. Other parameters are the same as in Fig. 2 (a) in the main text.}\label{Figbigws}
\end{figure}

Next, we discuss the condition for $T_f>\Lambda$. Here, we plot band structures when $T_f=2\Lambda$, $1.5\Lambda$ and $1.25\Lambda$ in Figs. \ref{figgap} (a) (c) and (e).
To analyze the impacts of periods, we fix the amplitude of $f(x)$. Note that $f(x)$ has period-dependent amplitude, and we need to change $a_0$ and $T_f$ simultaneously to make $a_0\alpha$ fixed. We choose $a_0\alpha=\frac{\pi}{4}$, which is the same as that of Fig. \ref{figsymband}(b). Therefore, the locations of DPs and slope of bands are the same in these figures but there are some new gaps in the band structures. In addition, the number of gaps increases from Fig. \ref{figgap} (a)-(e). That means that two different periods construct aperiodic structures with various orders. The robustness of locations of DPs are also the same as aperiodic GSLs \cite{Ma2012}. Then, we plot the electronic states versus different $T_f$ and find that with increasing $T_f$, the number of new gaps clearly increases, which means the order of aperiodic structures increases. These new gaps are also controllable by adjusting $T_f$. When $T_f$ is larger than a specific value, the energy bands become the discontinuous ones. Our results propose another way to construct aperiodic GSLs and acquire tunable band gaps by changing $T_f$. Meanwhile, curved graphene in experiments may not be exactly periodic \cite{Tapaszt2012}, so there may be some gaps in realistic CGSLs.

Previous studies reveal that new DPs that locate at $k_y\neq0$ could appear with some conditions in flat GSLs \cite{Park20081,Park2009,Brey2009,Barbier2010}. Now, we discuss the condition for the appearance of new DPs in CGSLs. We still consider the simplest situation $w_A=w_B$ and $T_f=\Lambda$ with the simplified model proposed above.

We also increase the periods of potentials to obtain new DPs. In Figs. \ref{Figbigws} (a)-(c), we plot band structures with $w_A=w_B=80$ nm and $a_0=0$, 40 nm and 50 nm. It is demonstrated that the slopes of band structures decrease such that they nearly vanish, and then, new DPs appear with increasing $a_0$. The $k_y$-direction coordinates of new DPs also increase with increasing $a_0$. These results illustrate that the space-dependent Fermi velocity makes the existence of new DPs easier. Thus, we plot electronic states with different $w_{A(B)}$ values. Comparing the results and those of Fig. 5(d) in Ref. \cite{Wang2010}, it indicates that new DPs can exist with smaller $w$ when $a_0\neq0$.

The condition for the existence of new DPs can also been discussed by the simplified model proposed above.
For new DPs, it has been illustrated in the literature \cite{Barbier2010,Wang2010,Fan2016} and Eq. (\ref{eqsimple}) that once the condition
\begin{equation}
q_Aw_A=-q_Bw_B=m\pi,\quad m=1,2,3...
\end{equation}
is satisfied under some specific $k_y$, $\sin(q_Aw_A)=\sin(q_Bw_B)=0$ and $\cos(q_Aw_A+q_Bw_B)=1$; then, $\cos(\beta_x\Lambda)=1$ and $\beta_x$ always has a real solution with all energies. This condition leads to the closing of the {\it zero-$\bar{k}$ gap}, and a pair of new DPs will appear away from $k_y=0$.
If we set $V_A=-V_B$ and $w_A=w_B$, then $f_A=f_B=f$, and the DPs should be located at zero energy,
so we discuss $E=0$ next, which refers to zero-energy modes studied in previous works\cite{Brey2009,Fan2016}.
Under the above assumption, we obtain
\begin{equation}
k_{y,m}=\pm\sqrt{(\frac{V}{\hbar v_f})^2-(\frac{2m\pi}{f\Lambda})^2},\quad m=1,2,3...,\label{eqnewDP}
\end{equation}
is satisfied with the above conditions.
The new DPs will exist when $(\frac{V}{\hbar v_f})^2-(\frac{2m\pi}{f\Lambda})>0$.
Therefore, when $f>1$, $k_{y,m}$ can obtain real solutions with smaller $\Lambda$ and can make the generation of new DPs easier.
This conclusion is consistent with Fig. \ref{Figbigws}(d).

%After setting $V_A=-V_B=2\pi l\frac{\hbar v_f}{\Lambda}$, %Eq.(\ref{eqnewDPV}) is obtained.
Eq. \ref{eqnewDP} can also be changed to
\begin{equation}
k_{y,m}=\pm\frac{2\pi}{\Lambda}\sqrt{l^2-(\frac{m}{f})^2},\quad m=1,2,3...,\label{eqnewDPV}
\end{equation}
with $V_A=-V_B=2\pi l\frac{\hbar v_f}{\Lambda}$. %, is the locations of DPs. New DPs with $k_y\neq0$ will exist when $l^2-(\frac{m}{f})^2>0$.
Here, $l$ represents potential since it is proportional to $V_A$.
In flat GSLs with $f=1$, a new pair of new DPs are generated once $l$ is a positive integer number larger than one \cite{Barbier2010,Fan2016}.
In CGSLs, $f>1$, so new DPs can arise with smaller $l$.

\begin{figure}[tb]
  \centering
  \includegraphics[width=8.5cm]{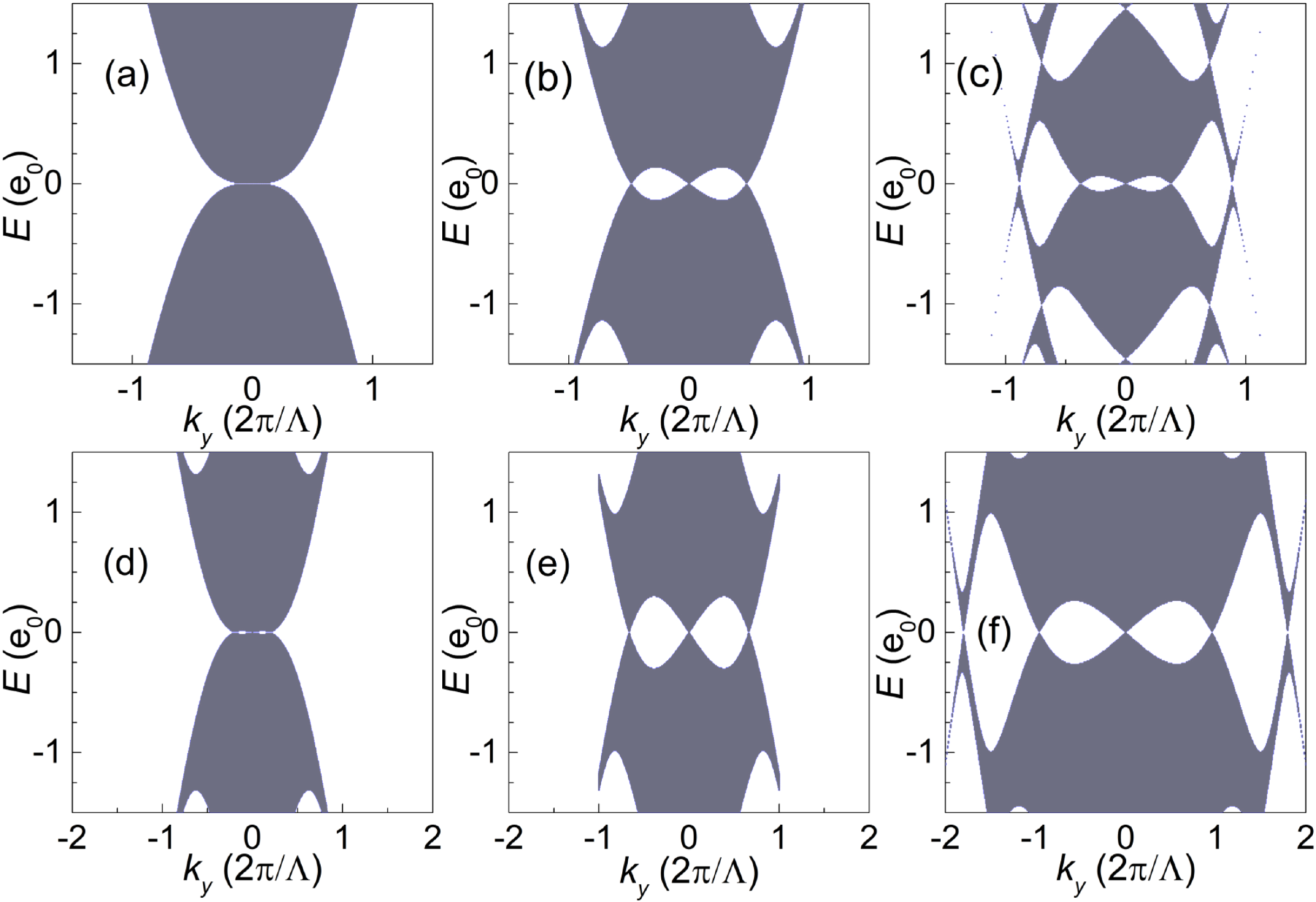}
  \caption{(Color online) (a)-(c) are electronic band structures with fixed $l=1$ and $a_0=0$, 15 nm and 55 nm, respectively. (d)-(f) are those with fixed $a_0=15$ nm and $l=0.9$, 1.1 and 2, respectively. Energy is in units of $e_0=\frac{\hbar v_f}{\Lambda}$. Other parameters are the same as in Fig. \ref{figsymband}(a).}\label{fignewDP}
\end{figure}

Fig. \ref{fignewDP} is plotted to verify the above conclusions. In Figs. \ref{fignewDP} (a)-(c), band structures with different $a_0$ and fixed $l=1$ are plotted. Here, we choose $a_0=0$, 15nm and 55nm to get clearly visible 1, 3 and 5 DPs, and we still use symmetric potential and $T_f=\Lambda$. It is illustrated that the slope of the band first decreases such that it nearly vanishes, and then, a new pair of DPs arise with increasing $a_0$. These phenomena are totally different from flat ones with $l=1$, where the first pair of new DPs just appear. Then, we set some $l$ and fixed $a_0=15$ nm to calculate the band structures in Figs. \ref{fignewDP} (d)-(f), showing the appearance of the first and second pair of new DPs. It is demonstrated that new DPs can generate in a smaller potential when graphene is curved. For example, there are only three DPs when $l=2$ in flat GSLs, but five in the CGSL with $a_0=15$ nm. In addition, the upper row of Fig. \ref{fignewDP} suggests that the locations of new DPs can be tuned by $a_0$ even with fixed potential or $l$, which cannot be realized in flat GSLs. Eq. (\ref{eqnewDPV}) also denotes this. By comparing the two rows of Fig. \ref{fignewDP}, one can also find that the coordinates of new DPs in upper rows are obviously smaller than those in the low row due to smaller $l$. Our discussions provide a possible method to adjust the locations of new DPs. Since the above effects originated from the change of the Fermi velocity, our results suggest that forming that periodic Fermi velocity by other ways \cite{Pellegrino2012,Raoux2010,Krstaji2011} may also be available.
%More discussions about new DPs can be found in SM IV.

\begin{figure}[tb]
  \centering
  \includegraphics[width=6.5cm]{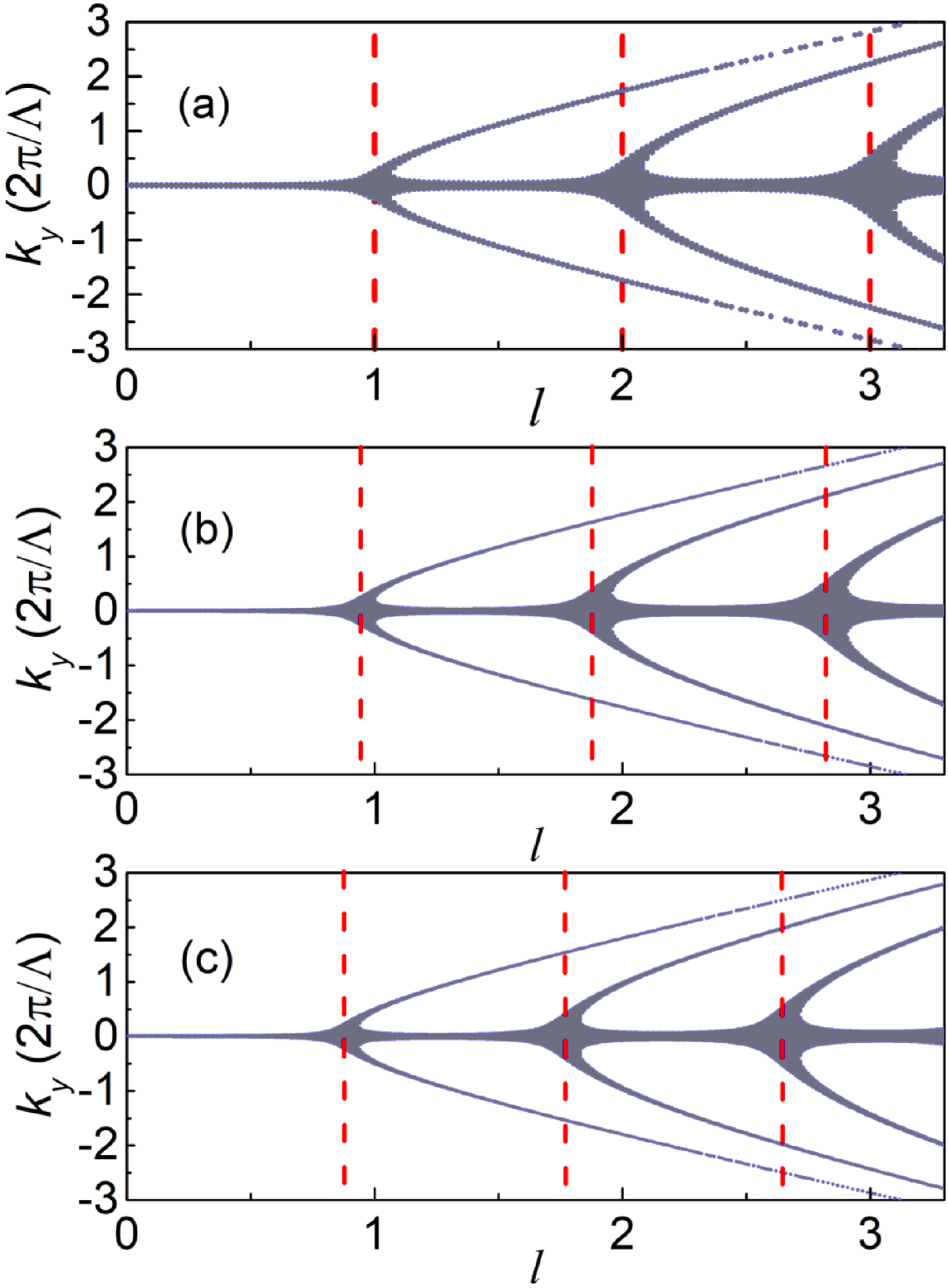}
  \caption{(Color online) Electronic states in different $l$ and $k_y$ with fixed $E=0$. Here, $a_0=0$ (a), $a_0=10$ nm (b) and $a_0=15$ nm (c) to see the change trend in a small area. Other parameters are the same as in Fig. \ref{Figbigws}. The red dashed lines denote the appearance of new DPs.}\label{FignumDPs}
\end{figure}

Finally, we discuss the number of DPs under different $l$ and $a_0$.
We compute the zero-energy electronic states with different $k_y$ and $l$ in Fig. \ref{FignumDPs}. When new DPs appear, states away from $k_y=0$ will be present in these figures.
Referring to the flat situation in Fig. \ref{FignumDPs}(a), we can indicate $l$ at which new DPs arise in (b) and (c). It is shown that $l$ under this condition is smaller with increasing $a_0$. The distinctions between the flat and curved situations also increase with larger $l$.
We can conclude that with space-dependent Fermi velocity induced by curvature, new DPs can arise with smaller potentials, which means this changed velocity works as an effective potential. In view of the fact that the Fermi velocity can be measured indirectly\cite{luican2011quantized,yan2013superlattice}, and there have been experiments that have observed the appearance of new Dirac points in quasi-periodic graphene\cite{yan2013superlattice}, our conclusion may be verified experimentally. This finding  is consistent with the analysis in the Results and discussions section and provides a possible simple way to acquire new DPs and then investigate properties near them.

We have discussed several situations in this part and here we made a list. (a). For the simplest situation with equal periods and symmetric potentials, the DPs will not move, but the slopes of bands decreased with increasing $a_0$. (b). For asymmetric potentials, there will be tunable shifted DPs which move to higher energy when $w_A>w_b$ and to lower energy when $w_A<w_B$. DPs shift more with increasing $a_0$. (c). For unequal periods, there will be extra gaps. So it is another way to construct aperiodic GSLs and acquire tunable band gaps by changing $T_f$. (d). By increasing the periods of potentials we can obtain new DPs and show that new DPs can exist with smaller $w$ in curvature. (e). We have also illustrated that new DPs can arise with smaller potentials by curvature.

\section{conclusions}
In summary, we proposed CGSLs by combining curved graphene and periodic potentials and then investigated their electronic properties. Since we focus on the effects of the space-dependent Fermi velocity, we use the Dirac equation in curved spacetime and the transfer matrix to obtain the band structures. For the simplest situation with equal periods and symmetric potentials, the DPs will not move, but the slopes of bands dramatically decrease in both the $k_x$- and $k_y$-direction. For asymmetric potentials and unequal periods, tunable shifted DPs and extra gaps can appear, respectively.
We also discussed the condition for DPs' appearance with symmetric potentials and equal periods. The condition for obtaining new DPs can be met more easily when graphene is curved. One can use this property to create new DPs experimentally to investigate physical phenomena near those points.

We would like to state that we reveal part of the space-dependent Fermi velocity's impacts on electronic properties. Although curvature can bring other effects such as the pseudomagnetic field, they are not reflected by our model and need to be analyzed by other means, such as the tight-binding model with elastic theory. Thus, it is valuable to discuss these effects with a more perfect method in the future. We would also like to emphasize that the effects of ripples in GSLs should not be neglected in some cases experimentally since they are intrinsic. That may cause a slight shift of DPs or extra gaps. The controllability of the DPs also provides a new way to adjust electronic structures in experiments.

%\noindent
%\underline{\it Acknowledgments}--
\begin{acknowledgments}
T.M. thanks CAEP for partial financial support. This work was supported by NSFC (Nos. 11774033, 11974049, 11674284 and 11974309), the Beijing Natural Science Foundation (No. 1192011) and the Zhejiang Provincial Natural Science Foundation of China under Grant No. LD18A040001. H.Q. Lin acknowledges financial support from NSAF U1930402 and NSFC 11734002, as well as computational resources from the Beijing Computational Science Research Center.
\end{acknowledgments}

\section*{Appendix A. Hamiltonian of curved graphene}
We rewrite the Dirac equation into a covariant form to obtain it in curved spacetime and the Hamiltonian with the method proposed by previous researchers \cite{Vozmediano2010}.
According to Eq. ({\color{blue}1}) %(\ref{eqflatH})
and the Dirac equation $i\hbar\frac{\partial}{\partial t}\psi=\hat{H}\psi$, we obtain its covariant form
\begin{equation}
i\hbar\bar{\gamma}^{\mu}\partial_{\mu}\psi=0,\label{App1}
\end{equation}
with $\mu=0,1,2$ representing time and $x$ and $y$ as the coordinates. We first use natural units $v_f=1$ during calculations. The short lines above $\bar{\gamma}^{\mu}$ illustrate the flat case and $\bar{\gamma}^{\mu}$ should satisfy the anticommutation relation $\{\bar{\gamma}^{\mu}, \bar{\gamma}^{\nu}\}=2\eta^{\mu\nu}I$ with the Minkowski metric $\eta_{\mu\nu}=\text{diag}(1, -1, -1)$. $\bar{\gamma}^{\mu}=(\sigma_3, -i\sigma_2, i\sigma_1)$ with $(\sigma_1, \sigma_2, \sigma_3)=(\sigma_x, \sigma_y, \sigma_z)$ as the Pauli matrices. These $\gamma$ matrices fit the above anticommutation relation.

Then, we rewrite Eq. (\ref{App1}) into the curved case \cite{Vozmediano2010}, which reads
\begin{equation}
i\hbar\gamma^{\mu}D_{\mu}\psi=0.\label{eqApp2}
\end{equation}
with metric $g_{\mu\nu}$. There are two differences between the curved and flat equations \cite{Vozmediano2010} (1) Fielbein fields $e^{\mu}_a$ need to be introduced to indicate the change of $\gamma^{\mu}$, which is $\gamma^{\mu}=e^{\mu}_a\bar{\gamma}^a$. Here, the symbols without short lines mean the curved case. Fielbein fields should fit conditions such that $g_{\mu\nu}=\eta_{ab}e_{\mu}^ae_{\nu}^b$,
$\gamma^{\mu}$ calculated from Fielbein fields should have $\{\gamma^{\mu}, \gamma^{\nu}\}=2g^{\mu\nu}I$ and the determinants of the metrics are $[\text{det}(g_{\mu\nu})]^{1/2}=\text{det}(e_{\mu}^a)$.
(2) The differential operator should be $D_{\mu}=\partial_{\mu}+\Omega_{\mu}$ and the spin connection is
\begin{equation}
\Omega_{\mu}=\frac{1}{4}e^{\nu a}(\partial_{\mu}e^b_{\nu}-\Gamma_{\mu\nu}^{\lambda}e^b_{\lambda})\bar{\gamma}_a\bar{\gamma}_b,\label{spinc}
\end{equation}
with the Christoffel symbol $\Gamma_{\mu\nu}^{\lambda}=\frac{1}{2}g^{\sigma\lambda}(\frac{\partial g_{\nu\sigma}}{\partial x^{\mu}}+\frac{\partial g_{\mu\sigma}}{\partial x^{\nu}}-\frac{\partial g_{\mu\nu}}{\partial x^{\sigma}})$.

In this paper, we consider a one-dimensional periodic curved surface, which refers to ripples that are only dependent on one coordinate and are written as $z=h(x)$. The line elements read
\begin{equation}
\begin{aligned}
ds^2&=dx^2+dy^2+dz^2=dx^2+dy^2+(\frac{dz}{dx})^2dx^2\\
&=(1+g^2(x))dx^2+dy^2,
\end{aligned}
\end{equation}
with $g(x)=\frac{dz}{dx}=h^{\prime}(x)$, so the metric is
\begin{equation}
g_{\mu\nu}=\text{diag}(1, -(1+g^2(x)), -1). \label{eqmetric}
\end{equation}
According to Eq. (\ref{eqmetric}), we obtain that
\begin{equation}
e_{\mu}^a=\text{diag}(1, \sqrt{1+g^2(x)}, 1),
\end{equation}
and $\Omega_{\mu}=0$.
After substituting them into Eq. (\ref{eqApp2}), multiplying $\sigma_z$ on both sides of the equation and adding $v_f$, we obtain Eq. (\ref{eqcurH}).

\section*{Appendix B. Transfer matrix for CGSL}
%Then, we calculate $H\Psi=E\Psi$ and obtain
After acting Eq. (\ref{eqSL}) on $\Psi=(\tilde{\psi}_A, \tilde{\psi}_B)^T=(\psi_A, \psi_B)^Te^{ik_yy}$, we can get
\begin{equation}
\begin{cases}
\frac{1}{f(x)}\frac{d}{dx}\psi_A-k_y\psi_A=ik\psi_B\\
\frac{1}{f(x)}\frac{d}{dx}\psi_B+k_y\psi_B=ik\psi_A
\end{cases}, \label{eqApp3}
\end{equation}
where $k=\frac{E-V(x)}{\hbar v_f}$ represents the wave vectors inside potentials. Due to the square barriers of potentials, $V(x)$ in the $m$th potential region maintains constant $V_m$.
Next, we divide the potential region into $n$ parts, so $f(x)$ changes little inside each part; then, $f(x)$ in the $j$th part can be regarded as constant $f_j$. With this approximation, Eq. (\ref{eqApp3}) is expressed as
\begin{equation}
\begin{cases}
\frac{d^2}{dx^2}\psi_A+f_j^2(k_j^2-k_y^2)\psi_A=0\\
\frac{d^2}{dx^2}\psi_B+f_j^2(k_j^2-k_y^2)\psi_B=0
\end{cases}.
\end{equation}
The following processes will be the same as in Ref. \cite{Wang2010}, and we can obtain Eq. (\ref{TM}).
\bibliography{Reference}

\end{document}